\documentclass[12pt]{article}
\usepackage{graphicx}
\usepackage[square,numbers]{natbib} 
\usepackage{url} 
\usepackage{amsmath, amssymb, amsthm}
\usepackage{graphicx,color}
\usepackage{multirow}
\usepackage[left=1in, right=1in, top=1in, bottom=1in]{geometry}
\usepackage{epsfig}
\usepackage{color}
\usepackage{setspace}
\usepackage{latexsym}
\usepackage{mathrsfs}
\usepackage{graphicx}
\usepackage{tikz}
\usepackage{listings}
\usepackage{algorithm}
\usepackage{subcaption}
\usepackage{algpseudocode}
\usepackage[titletoc]{appendix}
\usepackage{alltt}
\graphicspath{ {images/} }
\usepackage{longtable}
\usepackage{tikz}
\usepackage{acronym}
\usepackage{pgfplots}
\usepackage{shortvrb}
\usepackage{listings}
\usepackage{soul}
\usepackage{ulem}
\usepackage{booktabs}
\usepackage{csquotes}
\usepackage[hidelinks]{hyperref}
\usepackage{titletoc,tocloft}
\usepackage{bm}
\usepackage{leftidx}

\usepackage{times}
\usepackage{sectsty}
\usepackage{titlesec}

\newcommand{\blind}{0}

\begin{document}

\def\spacingset#1{\renewcommand{\baselinestretch}%
{#1}\small\normalsize} \spacingset{1}


\if0\blind
{
  \title{\bf O-MAGIC: Online Change-Point Detection for Dynamic Systems}
  \author{Yan Sun, Yeping Wang, Zhaohui Li, Shihao Yang \\
    Department of Industrial and Systems Engineering, Georgia Institute of Technology \\}
    \date{}
  \maketitle
} \fi

\if1\blind
{
  \bigskip
  \bigskip
  \bigskip
  \begin{center}
    {\LARGE\bf Title}
\end{center}
  \medskip
} \fi

\bigskip
\begin{abstract}

The capture of changes in dynamic systems, especially ordinary differential equations (ODEs), is an important and challenging task, with multiple applications in biomedical research and other scientific areas. This article proposes a fast and mathematically rigorous online method, called ODE-informed MAnifold-constrained Gaussian process Inference for Change point detection(O-MAGIC), to detect changes of parameters in the ODE system using noisy and sparse observation data. O-MAGIC imposes a Gaussian process prior to the time series of system components with a latent manifold constraint, induced by restricting the derivative process to satisfy ODE conditions.
To detect the parameter changes from the observation, we propose a procedure based on a two-sample generalized likelihood ratio (GLR) test that can detect multiple change points in the dynamic system automatically. 
O-MAGIC bypasses conventional numerical integration and achieves substantial savings in computation time. By incorporating the ODE structures through manifold constraints, O-MAGIC enjoys a significant advantage in detection delay, while following principled statistical construction under the Bayesian paradigm, which further enables it to handle systems with missing data or unobserved components. O-MAGIC can also be applied to general nonlinear systems. Simulation studies on three challenging examples: SEIRD model, Lotka-Volterra model and Lorenz model are provided to illustrate the robustness and efficiency of O-MAGIC, compared with numerical integration and other popular time-series-based change point detection benchmark methods.

\end{abstract}

\noindent%
\vfill

\newpage
\spacingset{2} 
\section{Introduction}
\label{sec:intro}

Change Point Detection (CPD) is a critical area of research in time series analysis, focusing on the identification of abrupt shifts in the statistical properties of sequential data\cite{xie2021sequentialquickestchangedetection}. Such changes often signify significant events that can adversely impact processes or systems, making timely detection essential for mitigating risks, minimizing losses, and informing prompt decision-making across various domains. In the context of dynamic systems modeled by Ordinary Differential Equations (ODEs), the importance of CPD becomes particularly interesting yet overlooked. ODEs are fundamental mathematical tools for representing and analyzing the behavior of dynamic systems in disciplines such as epidemiology \cite{li1995global}, ecology \cite{takeuchi2006evolution}, and physics \cite{pchelintsev2014numerical}. They capture the continuous-time evolution of system states, and the accuracy of these models heavily depends on accurate knowledge of key parameters, which are often not directly observable and must be inferred from empirical data. Detecting abrupt changes in these parameters -- effectively performing CPD within the framework of ODEs -- is crucial because such changes can indicate shifts in the underlying system dynamics that may not be immediately apparent from observational data alone. 

This problem is challenging for two primary reasons. First, detecting such changes for ODEs is more difficult compared to change detection for time series data. Due to the structure of ODEs, drastic changes in a specific component series may not necessarily indicate changes in parameters. Changes in the system state may not result in significant alterations in component series (see Sec \ref{sec:seird} for a more intuitive example). This distinction sets it apart from change point detection in time series data. Second, identifying these changes is more challenging than simply estimating constant parameters. The nonlinearity of ODEs and the discontinuity of parameters mean that the system outputs cannot be solved analytically. To solve the ODE given initial conditions and parameters, numerical integration methods are often required. Typical numerical methods, such as Euler's Method or Runge-Kutta Method \cite{lapidus1971numerical}, require extensive computation. Conventional methods for parameter change detection and parameter estimation involve repeated evaluation of the ODE solution, which is computationally expensive, especially for high complexity models with multiple parameters, further complicating the task of efficient CPD in ODEs and underscoring the need for advanced methodologies in this area. Despite its importance, this problem remains underexplored due to technical challenges mentioned above. Traditional methods often struggle with abrupt parameter shifts because they rely on smoothness assumptions that do not hold in such cases. In this paper, we aim to fill this gap by proposing a novel method that simultaneously detects parameter change points and estimates unknown parameters within the ODE framework. Our approach leverages Gaussian process inference with manifold constraints to efficiently handle discontinuities, thereby addressing a critical need in dynamic system analysis.

In this paper, we address these challenges by presenting a coherent method called ODE-informed MAnifold-constrained Gaussian process Inference for Change point detection (O-MAGIC), which can detect parameter changes and estimate unknown parameters in dynamic systems. Specifically, we assign a Gaussian process prior to system components. By incorporating the manifold constraint -- induced by restricting the derivative process to satisfy ODE conditions -- into the GP prior, the Bayesian posterior for unknown parameters and system components can be obtained. To detect change points in parameters, we propose a novel procedure based on a two-sample generalized likelihood ratio (GLR) test that can automatically detect multiple change points in the unknown parameter of the dynamic system.

The proposed method offers several key advantages:

\begin{enumerate}
\item Gaussian process enables the calculation of likelihood and posterior, making the GLR test for change point detection statistically principled;
\item The proposed method can detect change points in parameters and estimate the parameters before and after the change point, helping to quantify the extent of system change;
\item Compared to existing methods, our method avoids conventional numerical integration, making it more computationally efficient.
\end{enumerate}

As an efficient tool for estimating parameters in the system, O-MAGIC aims to locate parameter changes in ODE systems rather than directly analyzing observations on each dimension. We present three examples to demonstrate the effectiveness of O-MAGIC in detecting parameter changes in ODEs, where O-MAGIC outperforms competitive benchmark methods in accuracy and achieves superior computational efficiency compared to brute-force numerical methods.

The remainder of this paper is organized as follows. In Section \ref{sec:Literature review}, we provide a comprehensive literature review on relevant research areas, including studies on ODE systems and change point detection. In Section \ref{sec:meth}, we introduce the ODE-informed Gaussian process inference method, which serves as the foundation for our approach. Section \ref{sec:alg} presents the complete procedure of O-MAGIC. In Section \ref{examples}, we provide multiple examples to showcase the accuracy and computational efficiency of O-MAGIC. The conclusion of the paper is provided in Section \ref{discussion}, where sensitivity analysis, limitations, and future directions are discussed.

\section{Literature review}
\label{sec:Literature review}

Recent advances have led to the development of Bayesian inference methods utilizing Gaussian processes for parameter estimation in ordinary differential equations (ODEs) \cite{calderhead2008accelerating}, Partial Differential Equations (PDEs) \cite{chen2021solving}, and Stochastic Differential Equations (SDEs) \cite{pokern2013posterior, papaspiliopoulos2012nonparametric, hairer2011signal}. Several studies \cite{calderhead2008accelerating,dondelinger2013ode,barber2014gaussian,bera2017fast,lazarus2018multiphase,wenk2019fast} have applied Gaussian processes to facilitate ODE solving problems. More recent works have combined the advantages of Gaussian processes with Bayesian filtering methods \cite{tronarp2019probabilistic, kramer2022probabilistic} to solve initial value problems involved in ODEs. Gaussian processes have also been explored in inverse problems. \cite{yang2021inference} and \cite{calderhead2008accelerating} used Gaussian processes for constant and time-varying parameter inference. \cite{calderhead2008accelerating} utilized the product-of-expert heuristic to construct a GP posterior for time-constant parameter estimation. These methods offer computational efficiency and robust uncertainty quantification. 

However, a significant limitation is that they typically assume parameters are either constant or change smoothly over time. In many real-world systems, parameters can exhibit abrupt changes due to sudden events or interventions. For example, during the COVID-19 pandemic, policy changes or the implementation of public health measures can cause sudden shifts in the transmission rate $\beta$ \cite{biswas2014seir}. Accurately inferring whether $\beta$ has experienced significant changes within specific time frames, such as the past two weeks, is essential for effective response planning. Moreover, robust estimation of the transmission rate before and after such changes is crucial for understanding the dynamics of the outbreak.

Most change point detection studies analyze observations as independent multi-dimensional time series \cite{aminikhanghahi2017survey}, employing mainstream approaches such as CUSUM \cite{lee2003cusum} and auto-regression \cite{fryzlewicz2014multiple}. For instance, \cite{harchaoui2008kernel} proposed a change-point analysis for temporal observations based on a reproducing Hilbert kernel. \cite{dehling2014change} investigated the change of GARCH parameters in general time series based on CUSUM of squared residuals. \cite{kim2009test} studied tail index changes in stationary time series using the CUSUM test. Online change point detection for time series has also been researched \cite{aminikhanghahi2018real} \cite{agudelo2020bayesian}. 

Change detection of parameters has also been explored in SDEs. \cite{negri2012asymptotically} investigated a distribution-free statistic based on the Fisher-score process to detect drift parameter changes in diffusion processes. \cite{beibel1997sequential,dehling2014change} derived a GLR test for change detection of drift parameters in the general Ornstein-Uhlenbeck process. \cite{lee2006test} employed a CUSUM statistic based on a one-step estimator to detect parameter changes in diffusion processes.

Another line of research analyzes the subspaces in which time series sequences are constrained, with most studies using a modified Kalman filter to detect changes in a discrete-time linear state-space system online \cite{kawahara2007change, vellekoop2006nonlinear}, which do not possess the ability of detecting both change time and changed value from sparse and noisy data.

There are only a few papers focusing on modeling ODE parameters as piece-wise linear functions and thus identifying the change point. Most of them are designed for specific model structures, such as \cite{dehning2020inferring, jiang2020bayessmiles}, which employed Bayesian inference with Markov chain Monte Carlo (MCMC) to detect parameter changes in the Susceptible-Infected-Removed (SIR) model, and \cite{tykierko2007using}, which studied change detection for the Lorenz model based on model characteristics like the Hausdorff dimension. The challenge lies in generalizing these methods for a wider range of ODE models, as the literature on online parameter estimation and change detection for general non-linear ODEs is still scarce.

In summary, the existing literature on ODE parameter change detection is limited in common scenarios and often focuses on specific model structures. Our proposed O-MAGIC method aims to address these gaps by offering a computationally efficient and statistically principled approach to tackle parameter estimation and change detection in general non-linear ODE systems.

\section{ODE-Informed Manifold-Constrained Gaussian Processes}
\label{sec:meth}
Given the general formulation of ODEs that we are going to solve:
\begin{equation}
\bm{\dot x}(t) \equiv \frac{d\bm{x}(t)}{dt}=\mathbf{f}(\bm{x}(t),\bm{\theta}(t), \bm{\psi}, t), t \in [0,T]
\end{equation}
where $\bm{x}(t)$ represents the time series of system outputs from time $0$ to $T$. $\bm{\dot x}$ denotes the derivative of $\bm{x}$ with respect to time $t$; $\bm{\psi}$ denotes time-constant parameters; $\bm{\theta}(t)$ represents parameters modeled as a step function with multiple \textit{change points}; and $\mathbf{f}$ is a set of general functions characterizing the derivative process.

The ODE-Informed Manifold-Constrained Gaussian Processes(MAGI) \cite{yang2021inference} aims to construct a joint posterior distribution over model parameters $\bm{\theta}(t), \bm{\psi}$ and system states $\bm{x}(t)$ given sparse and noisy observations \( \bm{y}(\bm{\tau}) = \bm{x}(\bm{\tau}) + \bm{\epsilon}(\bm{\tau}) \). Which combines the likelihood function with prior distributions over the parameters and stochastic processes. 

Following MAGI paradigm, $\bm{x}(t)$ and time-varying parameters $\bm{\theta}(t)$ are realizations of $D$-dimensional and $P$-dimensional gaussian processes, other time-constant parameters is a realization of variable $\bm\Psi$. We specify a prior distribution \( \pi(\cdot) \) on \( \theta \) and independent GP priors on each component \( X_d(t) \), such that \( X_d(t) \sim \text{GP}(\mu_d, K_d) \), where \( K_d \) is a positive definite covariance kernel and \( \mu_d \) is the mean function. 

Observations are taken as \( \bm{y}(\bm{\tau}) = (y_1(\tau_1), \ldots, y_D(\tau_D)) \), with each component \( X_d \) observed at its own time points \( \tau_d = (\tau_{d,1}, \ldots, \tau_{d, N_d}) \), allowing for a flexible data collection scheme. In this context, $t$ generically represents time, while $\bm{\tau}$ specifically denotes observation time points. The symbol $\bm{I}$ signifies a finite discretization of time points $\bm{I} = \{t_1, t_2, \ldots, t_N\}$, with $|\bm{I}|=N$ and $\bm{\tau} \subset \bm{I} \subset[0,T]$.  

To capture the discrepancy in the derivative process $\bm{\dot X}(t)$ between the Gaussian process and the ODE, we employ the variable $W$ defined as $W=\mathop{\sup}\limits_{t \in [0,T],d \in \{1,...,D\}} \lvert \dot X_d(t)-\mathbf{f}(\bm{X}(t),\bm{\Theta}(t), \bm{\Psi}, t)_d \rvert$. $W=0$ holds if and only if $\bm{\dot X}(t)$ precisely conforms to the ODE structure, which is equivalent to constraining $\bm{X}(t)$ on the manifold of the ODE solutions. A computable approximation, denoted as $W_{\bm{I}}$, exists on the discrete set $\bm{I}$: $W_{\bm{I}}=\mathop{\sup}\limits_{t \in \bm{I},d \in \{1,...,D\}}\lvert \dot X_d(t)-\mathbf{f}(\bm{X}(t),\bm{\Theta}(t), \bm{\Psi}, t)_d \rvert$. The significance and advantages of $W$ are discussed in \cite{yang2021inference}.

\subsection{posterior}Intuitively, following the Bayesian paradigm, to detect the true time-varying parameters, we shall get the highest probability of true solutions, observation points, and the derivative satisfied with ODE equations given constant parameters, time-varying parameters and noisy level. i.e. 
\begin{equation}
\bm\Theta(\bm I) =
\underset{\bm\Theta(\bm I)}{\operatorname{argmax}} 
P_{W_{\bm I},\bm Y, \bm X | \bm \Theta, \bm \psi, \bm \sigma}    (W_{\bm I} = 0,     \bm Y(\bm \tau) = \bm y(\bm \tau), \bm X(\bm I)=\bm x(\bm I)  | \bm \Theta(\bm I) = \bm \theta(\bm I), \bm \Psi = \bm \psi, \bm \sigma)\label{eq:posterior} 
\end{equation}

\subsection{prior}
The prior distribution of $\bm{X}$ and $\bm\Theta$ in each dimension is independent Gaussian process: $X_d(t) \sim \mathcal{GP}(\mu_d^X, \mathcal{K}^X_d), t \in [0,T], d \in \{1,\ldots, D\}$, $\Theta_p(t) \sim \mathcal{GP}(\mu_p^\Theta, \mathcal{K}^\Theta_p), t \in [0,T], p \in \{1,\ldots, P\}$, where $\mathcal{K}^X_d$ and $\mathcal{K}^\Theta_p$: $\mathbb{R} \times \mathbb{R} \rightarrow \mathbb{R}$ is positive definite covariance kernel for GP, while $\mu_d^X$ and $\mu_p^\Theta$: $\mathbb{R} \rightarrow \mathbb{R}$ denote mean functions.

\subsection{likelihood}
With the help of GP and its analytical form of derivation. The goal here is to compute the posterior distribution of noisy observations denoted as $\bm{y}(\bm{\tau})=(y_1(\bm{\tau}_1),..., y_D(\bm{\tau}_D))$, where $\bm{\tau}=(\bm{\tau}_1, \bm{\tau}_2,...,\bm{\tau}_D)$ and each $\bm{\tau}_i \subset [0, T]$ consists of observation time points in $i^{th}$ component. We assume additive Gaussian observation noise: $Y_{d}(\bm{\tau}_{d}) = X_{d}(\bm{\tau}_{d}) + \epsilon(\bm{\tau}_d)$, with $\epsilon(\bm{\tau}_d) \sim \mathcal{N}(0, \sigma_{d}^2 )$. In this context, $t$ generically represents time, while $\bm{\tau}$ specifically denotes observation time points. The symbol $\bm{I}$ signifies a finite discretization of time points $\bm{I} = \{t_1, t_2, \ldots, t_N\}$, with $|\bm{I}|=N$ and $\bm{\tau} \subset \bm{I} \subset[0,T]$. We define the change points of parameter $\bm{\theta}(t)$ as $C = \{t \in [0,T]: \bm{\theta}(t^-) \neq \bm{\theta}(t^+)\}$. For simplicity, we assume all change points are contained within $\bm{I}$, i.e., $C\subset \bm{I}$, which is reasonable for a dense discretization set $\bm{I}$.

We first summarize several key components of our methodology:  
\begin{enumerate}
    \item The prior distribution  of $\bm{X}$ in each dimension is independent Gaussian process: $X_d(t) \sim \mathcal{GP}(\mu_d^X, \mathcal{K}^X_d), t \in [0,T], d \in \{1,\ldots, D\}$, where $\mathcal{K}^X_d$: $\mathbb{R} \times \mathbb{R} \rightarrow \mathbb{R}$ is positive definite covariance kernel for GP, while $\mu_d^X$: $\mathbb{R} \rightarrow \mathbb{R}$ denote mean functions.
    
    \item We employ a variable $W$ to capture the discrepancy in the derivative process $\bm{\dot X}(t)$ between the Gaussian process and the ODE: $W=\mathop{\sup}\limits_{t \in [0,T],d \in \{1,...,D\}} \lvert \dot X_d(t)-\mathbf{f}(\bm{X}(t),\bm{\Theta}(t), \bm{\Psi}, t)_d \rvert$. $W=0$ holds if and only if $\bm{\dot X}(t)$ precisely conforms to the ODE structure, which is equivalent to constraining $\bm{X}(t)$ on the manifold of the ODE solutions. A computable approximation, denoted as $W_{\bm{I}}$, exists on the discrete set $\bm{I}$: $W_{\bm{I}}=\mathop{\sup}\limits_{t \in \bm{I},d \in \{1,...,D\}}\lvert \dot X_d(t)-\mathbf{f}(\bm{X}(t),\bm{\Theta}(t), \bm{\Psi}, t)_d \rvert$. The significance and advantages of $W$ are discussed in \cite{yang2021inference}.

\end{enumerate}
With all the preliminaries above, the closed-form surrogate likelihood of the observations and and the latent system components can be derived as
\begin{small}
\begin{align}
&P_{W_{\bm I},\bm Y, \bm X | \bm \Theta, \bm \psi, \bm \sigma}
    (W_{\bm I} = 0, 
    \bm Y(\bm \tau) = \bm y(\bm \tau), \bm X(\bm I)=\bm x(\bm I)  | \bm \Theta(\bm I) = \bm \theta(\bm I), \bm \Psi = \bm \psi, \bm \sigma)\label{eq:full-posterior} \\
=& \underbrace{ P(\bm{X}(\bm{I})=\bm{x}(\bm{I})\rvert \bm{\Theta}(\bm{I})=\bm{\theta}(\bm{I}),\bm{\Psi}=\bm{\psi}) }_{\text{GP prior for ODE solution}} \notag \\
& \times \underbrace{ P(\bm{Y}(\bm{\tau})=\bm{y}(\bm{\tau})\rvert \bm{X}(\bm{I})=\bm{x}(\bm{I}), \bm{\Theta}(\bm{I})=\bm{\theta}(\bm{I}),\bm{\Psi}=\bm{\psi}, \bm \sigma) }_{\text{Observation Noise Likelihood}} \notag \\
&\times \underbrace{ P(W_{\bm{I}}=0\rvert \bm{Y}(\bm{\tau})=\bm{y}(\bm{\tau}),\bm{X}(\bm{I})=\bm{x}(\bm{I}),\bm{\Theta}(\bm{I})=\bm{\theta}(\bm{I}),\bm{\Psi}=\bm{\psi}, \bm \sigma) }_{\text{ODE-Informed manifold constraint}} \label{eq:decomposition} \\
=& \exp \Big\{ \sum^D_{d=1}\Big[ \rvert \bm{I}\rvert \log(2\pi)+\log\rvert \mathcal{K}^X_d(\bm{I})\rvert +\|x_d(\bm{I})-\mu_d^X(\bm{I})\|^2_{\mathcal{K}^X_d(\bm{I})^{-1}} \notag \\
&+N_d \log(2\pi\sigma_d^2)+\|x_d(\bm{\tau}_d)-y_d(\bm{\tau}_d)\|^2_{\sigma_d^{-2}}\notag\\
&+ \rvert \bm{I}\rvert \log(2\pi)+\log\rvert C_d\rvert +\|\mathbf{f}_{d,\bm{I}}^{\bm{x}, \bm{\theta}, \bm{\psi}} - \dot \mu_d^X(\bm{I})- \mathcal{'K}^X_d(\bm{I})\mathcal{K}^X_d(\bm{I})^{-1} \{x_d(\bm{I})-\mu_d^X(\bm{I})\}\|^2_{C_d^{-1}}\Big] \Big) \Big\}, \label{eq:posterior-calculation}
\end{align}
\end{small}
where  $\|v\|_A^2=v^T Av$, $\rvert \bm{I}\rvert $ is the cardinality of $\bm{I}$, and $\mathbf{f}_{d, \bm{I}}^{\bm{x}, \bm{\theta}, \bm{\psi}}$ is short for the $d$-th component of $\mathbf{f}(\bm{x}(\bm{I}), \bm{\theta}(\bm{I}), \bm{\psi}, t_{\bm{I}})$, and $C_d = \mathcal{K''}^X_d(\bm{I}) - \mathcal{'K}^X_d(\bm{I})\mathcal{K}^X_d(\bm{I})^{-1}\mathcal{K'}^X_d(\bm{I})$ is the conditional covariance matrix of $\dot X_d(\bm{I})$ given $X_d(\bm{I})$. We choose Matern kernel with the degree of freedom $\nu=2.01$ for $\bm{X}(t)$ to guarantee a differentiable GP that allows more flexible patterns: 
\begin{equation}\label{eq:kernel}
\mathcal{K}_\nu(l)=\phi_1^2 \frac{2^{1-\nu}}{\Gamma(\nu)} (\sqrt{2\nu} \frac{l} {\phi_2})^\nu K_\nu(\sqrt{2\nu} \frac l \phi_2), \quad l=\rvert s-t \rvert 
\end{equation}
where $K_\nu$ denotes the modified Bessel function of the second kind. In this case, $'\mathcal{K}=\frac{\partial}{\partial s}\mathcal{K}(s,t)$, $\mathcal{K}'=\frac{\partial}{\partial t}\mathcal{K}(s,t)$, and $\mathcal{K}''=\frac{\partial^2}{\partial s \partial t}\mathcal{K}(s,t)$ are all well-defined.


\section{Parameter Change Detection}\label{sec:alg}

This section introduces a novel method for the real-time detection of parameter shifts in nonlinear ODE systems. Section 4.1 delineates both our online and offline change point detection strategies. In Section 4.2, we articulate an algorithm for evaluating the uncertainty associated with the locations of parameter changes. Finally, Section 4.3 provides a summary of the proposed procedure.

\subsection{Online Procedure via Generalized Likelihood Ratio Test}
This section elaborates on the comprehensive computational framework for online change point detection. The implementation is available on GitHub for reference. Commencing with an appropriate initialization (Subsection 4.1.1), we conduct a sequence of operations: hypothesis testing (Subsection 4.2.1), followed by hyperparameter re-evaluation (Subsections 4.2.2 and 4.2.3) upon each update of the observations.

\subsubsection{Initialization of Online Procedure}\label{online-1}
The online methodology proposed for change point detection entails the repeated execution of the generalized likelihood ratio (GLR) test in accordance with each successive time window update. 

Initially, we establish a dynamic sliding time window, denoted as $\mathcal T$, also referred to as the scanning window \cite{yau2016inference}. The scanning radius (or window length) is set to $T_{max}$, thus initializing $\mathcal T$ as $\mathcal{T}_0 = [0,T_{max}]$. Following \cite{yau2016inference}, we assume that with a suitably chosen scanning radius $T_{max}$, any window would contain at most one change point. 

Moreover, we posit that there are no change points within a brief initial period of the ODE systems. All parameters $\bm \theta$, $\bm \psi$, and kernel hyperparameters $\phi$ are estimated using observations $\bm y(\bm \tau)$ within the initial window $\mathcal{T}_0$. Given these assumptions, $\bm \theta$ is reduced to constant parameters in $\mathcal{T}_0$, allowing for a straightforward estimation procedure using existing approaches \cite{sun2021manifold}. 

Following the completion of these initial settings, procedures \ref{online-2} to \ref{online-4} are executed when new observations arrive. That is, with the $i^{th}$ arrival, the observation at time $\tau_i$ becomes available, and the observations within the current scanning window $\bm y(\bm \tau)=\bm y_{\tau_{i-N_i}:\tau_{i}}$ are active for change-point detection.

\subsection{Determination of empirical threshold}\label{threshold} 
Calculation of $h$ is often challenging, which involves approximation of distribution and asymptotic analysis \cite{yau2016inference}. However, the formulation of the problem-specific likelihood is complicated and varies between different models. Therefore, rigorous mathematical calculations are hard for this problem. 

Thus we propose a method to acquire an empirical threshold using the initial conditions and parameters estimated from the start of the series, we generate an artificial observation series $y'(t)$. By applying the GLR test mentioned in Section \ref{online-2}, we obtain a series of likelihood ratios under the null hypothesis and take the maximum value as the critical value of the model threshold. Experiments have confirmed that this approach helps to identify a proper $h$ as the threshold and that the performance is robust to the choice of $h$. 

\subsubsection{Generalized Likelihood Ratio Test}\label{online-2}

In our approach, the change point detection is cast as a hypothesis testing within a chosen scanning window. In particular, we test whether a change point $t_k$ exists within the current series:

\begin{align*}
    H_0&:\bm \theta_{t_{i-N_i}} =...= \bm \theta_{t_{i-1}} =...=\bm \theta_{t_{i}} \\
    H_1&: \exists k \in \{i-N_i,..., i\},\bm \theta_{t_{i-N_i}} =...= \bm \theta_{t_{k}} \neq \bm \theta_{t_{k+1}} = ...=\bm \theta_{t_{i}} 
\end{align*}

We denote the parameter space under the null hypothesis as $\bm \Theta_0$ and calculate the maximum likelihood assuming $\bm \theta$ is constant within the window:

\begin{align*}
    L_0=\max \limits_{\bm\theta \in \bm \Theta_0,\bm x,\bm \sigma,\bm \psi}
    P_{W_{\bm I},\bm Y, \bm X | \bm \Theta_0, \bm \psi, \bm \sigma}
    .
\end{align*}
Under the alternative hypothesis that $\bm \theta$ has one change point in the sliding window, the alternative parameter space expands to the set of all two-piecewise constant functions, denoted as $\bm \Theta_k$, with the break point at time $t_k$. To optimize computation, we limit $k$ to the detection zone of final $R$ observations (where $R \leq N_{i}$), identifying the optimal $\bm \theta \in \bm \Theta_{k}$ that yields the maximum likelihood:

\begin{align}
    &L_1=\max \limits_{k \in \{i-R,...,i\}}  \max \limits_{\bm \theta \in \bm \Theta_k,\bm x,\bm \sigma,\bm \psi} 
    P_{W_{\bm I},\bm Y, \bm X | \bm \Theta_k, \bm \psi, \bm \sigma}. 
\end{align}
The generalized likelihood ration test is implemented by calculating the test statistic $\log \Lambda =\log  \frac{L_1}{L_0}$ as the log-likelihood-ratio. If $\log \Lambda$ exceeds the pre-determined threshold $h$, the null hypothesis is rejected, signifying a change point within the window. The change point's inferred time point is thus $k^* = \arg\max \limits_{k} L_1$. otherwise, we conclude that no change point is detected within the window, and update the parameters $\bm \theta$ and $\bm \Psi$. 

\subsubsection{Updating the Sliding Window}\label{online-3}

The sliding window, denoted as $\mathcal T$ and previously introduced in \ref{online-1}, gets updated every time a new observation is incorporated. If we detect a change point within the current sliding window, we initiate a new $\mathcal T$ beginning immediately after the detected change point. This new window, which begins right after the change point, starts at the smallest allowable window size $T_{min}$ (usually set at $T_{min}=2$). The parameters from the previous iteration are carried over to this new window. If no change point is detected, we simply extend the current time window $\mathcal T$ by adding the new data. This extension continues until the window's size reaches the predefined maximum size, $T_{max}$. At this point, the oldest observation is discarded as the window transitions into a sliding mode.

\subsubsection{Update Kernel Hyperparameters}\label{online-4}

Due to the structure of ODE, when the scale of $\bm y(\bm \tau)$ is drastically different from initial observations in range and size, re-evaluation of the Gaussian kernel is recommended. This is achieved by maximizing the marginal likelihood of $\bm y(\bm \tau)$ as a function of Matern kernel hyperparameters $\phi_1$ and $\phi_2$:
\begin{align}\label{hyper-update}
\hat{\phi}_{1,d}, \hat{\phi}_{2,d} =\arg\max\limits_{\phi_1,\phi_2} \pi_{\bm{\Phi}_p}(\bm{\phi})P(\bm y(\bm \tau)^{(d)}| \bm{\phi}),  d=1,2,..., D.
\end{align}
A more cost-efficient approach is to optimize only $\hat{\phi}_{1,p}$ in Eq.\ref{hyper-update}, while holding $\hat{\phi}_{2,p}$ identical as the initial setting if no change point is detected in the sliding window. We denote the updated kernel hyperparameters in $i^{th}$ iteration as $\hat \phi^{(i)}$, and return to step \ref{online-2}.

\subsection{Uncertainty Quantification for Parameter Changes Identification}\label{sec:uncertainty}
In this section, we further build a Bayesian method on top of the likelihood equation Eq.\eqref{eq:posterior-calculation} for quantifying the probability that each time point has parameter changes. The key idea is to impose spike-and-slab prior to $\theta(t)$ \cite{ishwaran2005spike}. The ``spike'' part of the prior restricts $\theta(t_i)$ to be close to $\theta(t_{i-1})$, while the ``slab'' part of the prior allows $\theta(t_i)$ to take any value unrelated to $\theta(t_{i-1})$. The uncertainty quantification is achieved through posterior samples of $\bm \theta(t)$ using a Gibbs sampler.

\subsubsection{Stochastic process prior for $\theta(t)$} \label{sec:4.2.1}
In plain language, the Stochastic process prior is a slowly-drifted Brownian motion, superimposed with a Poisson arrival of change point. To write out the math, we restrict ourselves on discretized time points $\bm{I} = \{t_i\}$. We first define an auxiliary variable $A_i$, where $A_i=1$ when $\theta(t_i)$ has a parameter change from $\theta(t_{i-1})$, and $A_i=0$ otherwise.



With the introduction of $A_i$, we are ready to write out the spike-and-slab prior for $\theta(\bm{I})$. When $A_i = 0$, $\theta(t_{i}) | \theta(t_{i-1}), A_i = 0$ follows spike prior of a slowly-evolved Brownian motion $\mathcal{N}(\theta(t_{i-1}), \sigma_0^2 (t_{i} - t_{i-1}))$. When $A_i = 1$, $\theta(t_{i}) | \theta(t_{i-1}), A_i = 1$ follows slab prior of uniform distribution on a pre-set interval $U[\bm \theta_{min}, \bm \theta_{max}]$. Finally, the $A_i$ distribution is characterized by a Poisson process with rate $\lambda$, which gives the approximate independent Bernoulli prior $P(A_{t_{i}}=0) = e^{-\lambda_0 (t_{i}-t_{i-1})}$.


\subsubsection{\textbf{The posterior}}
With the $\theta(t)$ prior above characterized by $(\theta(I), A)$, we can derive the $\theta(t)$ posterior from the likelihood in Eq.\eqref{eq:full-posterior} as

\begin{align}\label{cpd:posterior}
&  P_{\bm{A}, \bm{\Theta}(\bm{I}), \bm{X}(\bm{I})|W_{\bm{I}}, \bm{Y}(\bm{\tau})}(\bm{A}, \bm{\theta}(\bm{I}), \bm{x}(\bm{I})|W_{\bm{I}}=0, \bm{Y}(\bm{\tau})=\bm{y}(\bm{\tau})) \notag\\
\propto& \underbrace{\pi_{\bm{A}, \bm{\Theta}(\bm{I})}}_{\text{parameter prior}}  \times \underbrace{P_{W_{\bm I},\bm Y, \bm X | \bm \Theta, \bm \psi, \bm \sigma}
    (W_{\bm I} = 0, 
    \bm Y(\bm \tau) = \bm y(\bm \tau), \bm X(\bm I)=\bm x(\bm I)  | \bm \Theta(\bm I) = \bm \theta(\bm I), \bm \Psi = \bm \psi, \bm \sigma)}_{\text{likelihood part (refer to Eq.\eqref{eq:full-posterior})}}
\end{align}
where the parameter prior can be written as
\begin{align}\label{mcmc:prior}
&\pi_{\bm{A}, \bm{\Theta}(\bm{I})}(\bm{A}, \bm{\theta}(\bm{I})) \notag \\
=&P(\bm A) \times \prod^{N}_{i=1}[P(\bm\theta(t_i)|\bm\theta(t_{i-1})), \bm A] \notag \\
=\Big[&\prod_{i \in \{i:A_i=0\}}e^{-\lambda_0 (t_{i}-t_{i-1})} \cdot \prod_{j \in \{j:A_j=1\}}(1-e^{-\lambda_0 (t_{j}-t_{j-1})})\Big] \times \notag \\
&\prod_{p=1}^P \Big[(\frac{1}{\theta_{max}^{(p)}-\theta_{min}^{(p)}})^{||\bm A ||_1^1}\cdot \prod_{i \in \{i:A_i=0\}} \frac{1}{\sqrt{2\pi} \sigma_0^{(p)}}\exp\{\frac{(\theta(t_i)^{(p)}-\theta(t_{i-1})^{(p)})^2)}{2 (\sigma_0^{(p)})^2(t_i - t_{i-1})}\}\Big]
\end{align}

\subsubsection{Sampling Scheme}
First, we initialize $\bm A$ at $\bm A^{(0)}$.
For example, all points are change points corresponding to $\bm A^{(0)} = (1,1,\dots,1)$. Then, we use systematic scan Gibbs sampling to update $A$. We propose a new sample $\bm A$ by flipping it at location $i$ sequentially for each dimension in $\bm A$ and then propose $\theta^{new}$ and $x^{new}$. We accept/reject the flip of $A_i$ according to the posterior ratio of new samples and current samples. Lastly, repeat the above procedure until convergence.

\subsection{Summary}

In this section, we provide an online method to detect parameter changes and an offline method to evaluate the probability of parameter change for each time point. The two approaches combined provide an accurate and robust framework for the evaluation of general ODE systems. In the online algorithm, we apply a sequential general likelihood test based on the empirical threshold when the scanning window is updated, and dynamically tune the window size and kernel hyperparameters. In the offline uncertainty quantification algorithm, we model the parameters as concatenated Brownian motion and apply Gibbs sampling method to sample the posterior distribution of parameter changes.

\section{Numerical Studies}\label{examples}

In this section, we assess the effectiveness of our method in identifying change points in an online setting using three famous models i.e. LV model, Lorenz and Susceptible-Exposed-Infectious-Recovered-Deceased (SEIRD) model. The results demonstrate that our proposed method consistently outperforms reasonably well in all these scenarios.


\textbf{Evaluation metrics:}
We employ several evaluation metrics commonly used in change point detection to showcase the 
 satisfactory performance of our methods, referencing the work of \cite{aminikhanghahi2017survey}. 
\begin{itemize}
\item When a change point is detected whereas no change point exists in the neighborhood, a false alarm is raised. The \textbf{false alarm rate (FAR)} is defined as: $ \textbf{FAR} = \frac{\text{\# False positive}}{\text{\# False positive} + N - \text{\# True changes}}$, FAR is a.
\item Given a change point is correctly detected, \textbf{expected detection delay (EDD)} quantifies the time gap between the detection moment and the actual occurrence of a change: $\textbf{EDD} = E(C^{\text{detect}}-C)$.
\item With our detection window, our algorithm estimates change points to be earlier than the detection time. Although a true change point must occur before its detection, the estimated change point may be earlier. Therefore, we use \textbf{mean absolute error (MAE)} to measure the time difference between the estimated and true change point: $\textbf{MAE} = E(|\hat C^{\text{est}}-C|)$.
\item Given the problem context, change point detection involving multiple change points inherently seeks to segment the time series into distinct regions. As such, employing clustering metrics such as the variation of information \cite{arabie1973multidimensional}, adjusted Rand index \cite{hubert1985comparing}, Hausdorff distance \cite{hausdorff1927mengenlehre}, and segmentation covering metric \cite{arbelaez2010contour,everingham2010pascal}, among others, might be the most effective evaluation approach. In our experiments, we opt for the covering metric \cite{van2020evaluation} as our clustering metric of choice.

For two subsets $\mathcal A, \mathcal A' \subseteq [1, T]$, the Jaccard Index, also known as Intersection over Union, is given by $J(\mathcal A, \mathcal A')=\frac{|\mathcal A \cap \mathcal A'|}{|\mathcal A \cup \mathcal A'|}$.
Following Arbelaez et al. (2010), we define the \textbf{covering metric} of a partition $\mathcal G$ by a
partition $\mathcal G'$ as $C(\mathcal G, \mathcal G')=\frac 1T \sum_{\mathcal A \in \mathcal{G}}|\mathcal A|\cdot \max\limits_{\mathcal A' \in \mathcal{G'}} J(\mathcal A, \mathcal A')$.
For a collection $\{\mathcal G_k\}_{k=1}^K$ of ground truth partitions and a partition $\mathcal{S}$ given by an algorithm, we compute the average of $C(\mathcal G_k, \mathcal S)$. 
\end{itemize}

Through the simulation of three well-known examples, we illustrate the computational efficiency of our approach compared to the brute-force method, and showcase the significant benefit of integrating ODE structure information compared to other general time series change-point detection approaches. 100 replications with change points randomly sampled from the same Poisson distribution are conducted in each experiment.
\subsection{Susceptible-Exposed-Infectious-Recovered-Deceased Model}\label{sec:seird}
Consider a hypothetical disease cases/deaths modeling using an infectious disease Susceptible-Exposed-Infectious-Recovered-Deceased (SEIRD) compartmental ODE model \cite{hethcote2000mathematics, hao2020reconstruction}, where an entire population is classified into S, E, I, R, D components, and any transitions from one state to another, representing the disease spread dynamics, are modeled as an ODE:
\begin{equation*}
\frac{dS}{dt}=-\frac{\beta IS}{N}, \quad
\frac{dE}{dt}=\frac{\beta IS}{N} - v^e E,  \quad
\frac{dI}{dt}=v^e E - v^i I, \quad
\frac{dD}{dt}=v^i I \cdot p^d,
\end{equation*}
where $N$ is the total population, and the cumulative recovered population is $R = N - S -E -I-D$. The $S, E, I$, and $D$ denote the susceptible, exposed, infected population and cumulative death respectively. 4 parameters of interest are investigated: rate of contact by an infectious individual ($\beta$), rate of transferring from the state of exposure to infectious ($v^e$), rate of leaving infectious period ($v^i$), and fatality rate ($p^d$). During a pandemic, parameters in the SEIRD model may contain abrupt changes due to pharmaceutical and non-pharmaceutical interventions. We assume that $\beta$ series contains multiple change points due to public policy interventions during a specific time, such as city lockdown; $p^d$ may also contain change points depending on the sufficiency of medical treatments and mutation. When the medical resources are drained due to the pandemic, $p^d$ will see a prompt increase. $v^e$ and $v^i$ are assumed to be unknown constants throughout the entire time horizon.

To fully emulate a realistic scenario, in the experiment we set
\begin{align*}
    \beta(t)=\left\{
\begin{aligned}
0.8, t < t_0 \\
0.1, t \geq t_0 \\
\end{aligned}
\right., p^d(t)=\left\{
\begin{aligned}
0.02, t < t_1 \\
0.05, t \geq t_1 \\
\end{aligned}
\right.
\end{align*}
where $t_0 \sim U[50, 70]$ denotes the time of public health policy changes, which leads to changes in $\beta$. $t_1 \sim U[90, 110]$ denotes the time of death rate change, due to the collapse of the healthcare system. The parameter values are set as $v^e=0.1, v^i=0.1$, and initial states are set as $[S_0, E_0, I_0, D_0]=[1000000, 1000, 500, 50]$. A new observation is acquired at a daily frequency with 5\% multiplicative noise. The system is analyzed by taking the log of each component. In the experiment, the detection zone $R$ is set as 7 days, and a moving window of size 40 is chosen in the online detection approach.

\begin{figure}[htp]
     \centering
     \begin{subfigure}[b]{1\textwidth}
         \centering
         \includegraphics[width=4in]{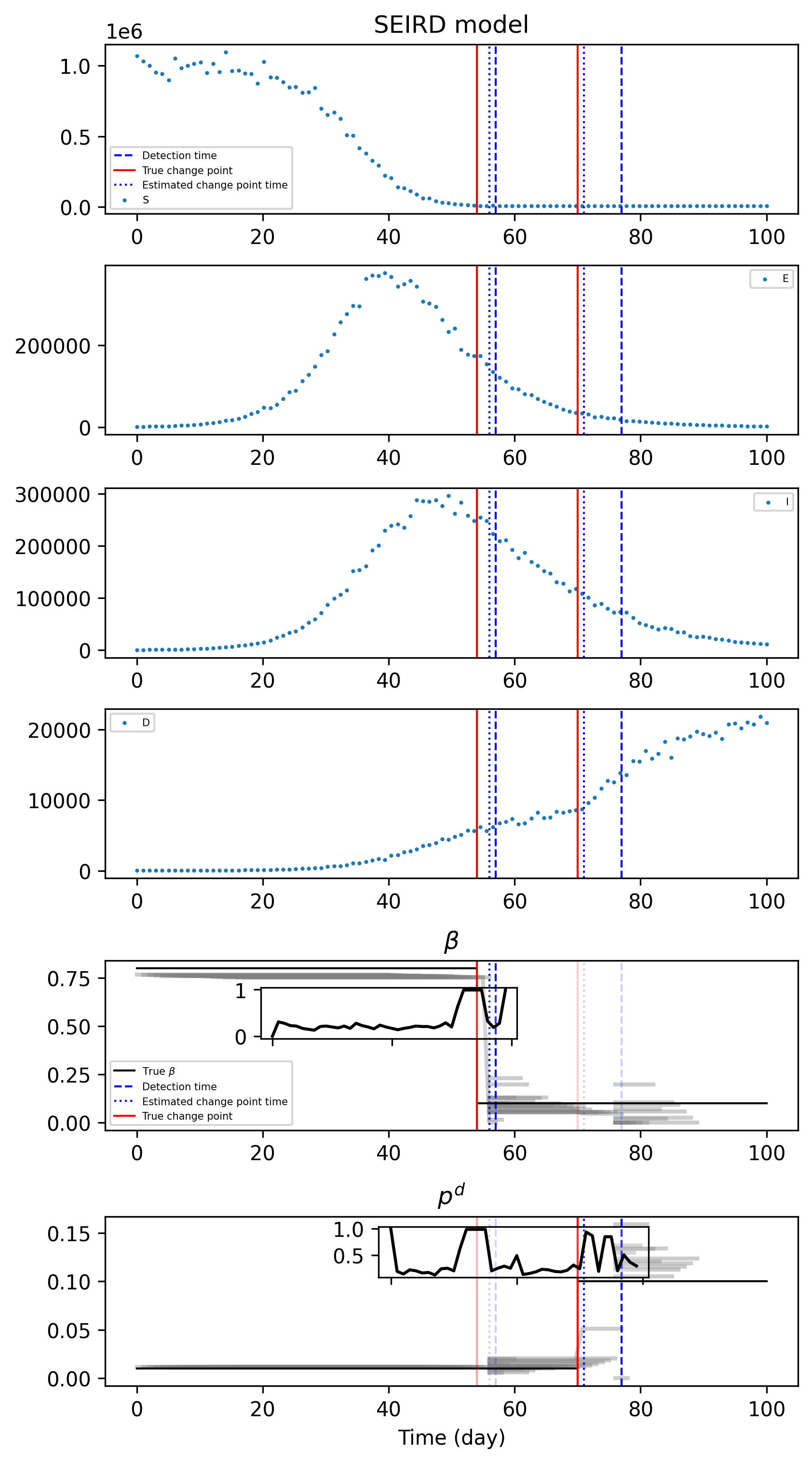}
     \end{subfigure}
\caption{Results of SEIRD model given one random sample observation. The top 4 plots show the sample observation with noise, where solid red lines indicate actual change time, blue dash lines indicate delayed detection time, and blue dotted lines denote the estimated change point. Two plots at the bottom show the ground truth parameters (in black)and estimated parameters (in gray). The embedded subplots show the probability of change occurrence for each time point.}\label{fig:seird}
\end{figure}

Figure \ref{fig:seird} illustrates the result of O-MAGIC in one random replication, which shows its strong ability in estimating the changes on multi-dimensional parameters. The Expected Detection Delay (EDD) of the detection time is fewer than 6 days, and the retrospectively estimated change time is even closer to the ground truth. Table \ref{tab:seird-1} shows the comparison of results and computing time. 

\begin{table}[htp]
\begin{center}
\scalebox{0.7}{
\begin{tabular}{ccccccc}
\toprule
Model name & FAR (\%) & EDD & MAE & Cover & Computation time (min) \\
\midrule
O-MAGIC & 4.82 $\pm$ 1.90 & 5.85 $\pm$ 1.05 & 1.98 $\pm$ 0.45 & 0.94 $\pm$ 0.12 & 38.5 $\pm$ 5.1 \\
Runge-Kutta & 3.55 $\pm$ 1.66 & 4.80 $\pm$ 1.61 & 2.14 $\pm$ 0.68 & 0.98 $\pm$ 0.16 & 79.9 $\pm$ 9.4 \\
Microsoft SSA & 21.47 $\pm$ 5.53 & 18.81 $\pm$ 11.36 & 18.81 $\pm$ 11.36 & 0.61 $\pm$ 0.23 & 1.8 $\pm$ 0.2 \\
TIRE & 28.33  $\pm$ 6.62 & 11.37 $\pm$ 3.62 &  11.37 $\pm$ 3.62 & 0.69 $\pm$ 0.24 & 2.9 $\pm$ 0.2 \\
\bottomrule
\end{tabular}
}
\end{center}
\caption{The mean of each evaluation metric and computation time is reported first with the standard deviation across 100 replications followed after $\pm$ sign.}
\label{tab:seird-1}
\end{table}

We see that compared to benchmark approaches, O-MAGIC can accurately locate the changes within the least detection delay. Moreover, it yields accurate estimates of the unknown parameters after the change, which is crucial in evaluating the efficacy of lock down policies, as well as estimating medical shortage. 

The quantification of uncertainty of changes is also investigated. The following $\theta(t)$ prior settings are adopted in the experiment: $\sigma_0=0.01, [\beta_{\min},\beta_{\max}] =[0, 1], [p^d_{\min},p^d_{\max}]=[0, 1], \lambda_0=1$. 1000 posterior samples of $\bm A$ are drawn after discarding the first 500 burn-in samples, following the algorithm described in Supplementary Material \ref{alg:hmc}, and the posterior mean of $\bm A$ is shown in Figure \ref{fig:seird} subplots. From the results, there is a significant peak near the truth change point, indicating a strong alert signal around the actual timestamp. 

\subsection{Lotka-Volterra Model}

Lotka-Volterra(LV) model (a.k.a. predator-prey model) is widely used to describe population fluctuation of predators, preys, and their interactions in the ecosystem \cite{goel1971volterra}. Due to the natural and man-made impact on the ecosystem, the interaction behaviors between species, such as the birth rate of a species, may suffer from abrupt changes. Capturing such changes is helpful for ecologists when monitoring the ecosystem and deciding possible interventions. However, the task is challenging due to the intricate interactions between species and the model's inherent seasonality, making it difficult to spot these changes relying solely on the usual time-series-based methods.

Specifically, the ODE system is characterized as:
\begin{equation*}
\frac{dx}{dt}=\alpha_t x - \beta xy, \quad
\frac{dy}{dt}=\delta xy - \gamma_t y,
\end{equation*}
where $x$ and $y$ denote the population of preys and predators. $\gamma_t$ denotes the death rate of the predator and is assumed to contain potential changes. $\alpha_t$ indicates the birth rate of the prey, $\beta$ and $\delta$ describe the interaction relationships between predators and preys. All parameters are assumed constant but unknown. We set the parameters $\beta=0.75$, $\delta=1, \alpha=0.6$. The birth rate of prey switches between $0.6$ and $1$ randomly with a Poisson process rate of 0.08. The time is measured on a monthly basis, and 1000 data point is obtained contaminated by 5\% multiplicative log-normal noise. The initial values of predators and preys are 1 and 2, as an ideal ratio in real ecology systems \cite{donald2003resistance}. 

\begin{figure}[htp]
     \centering
     \begin{subfigure}[b]{1\textwidth}
         \centering
         \includegraphics[width=4in]{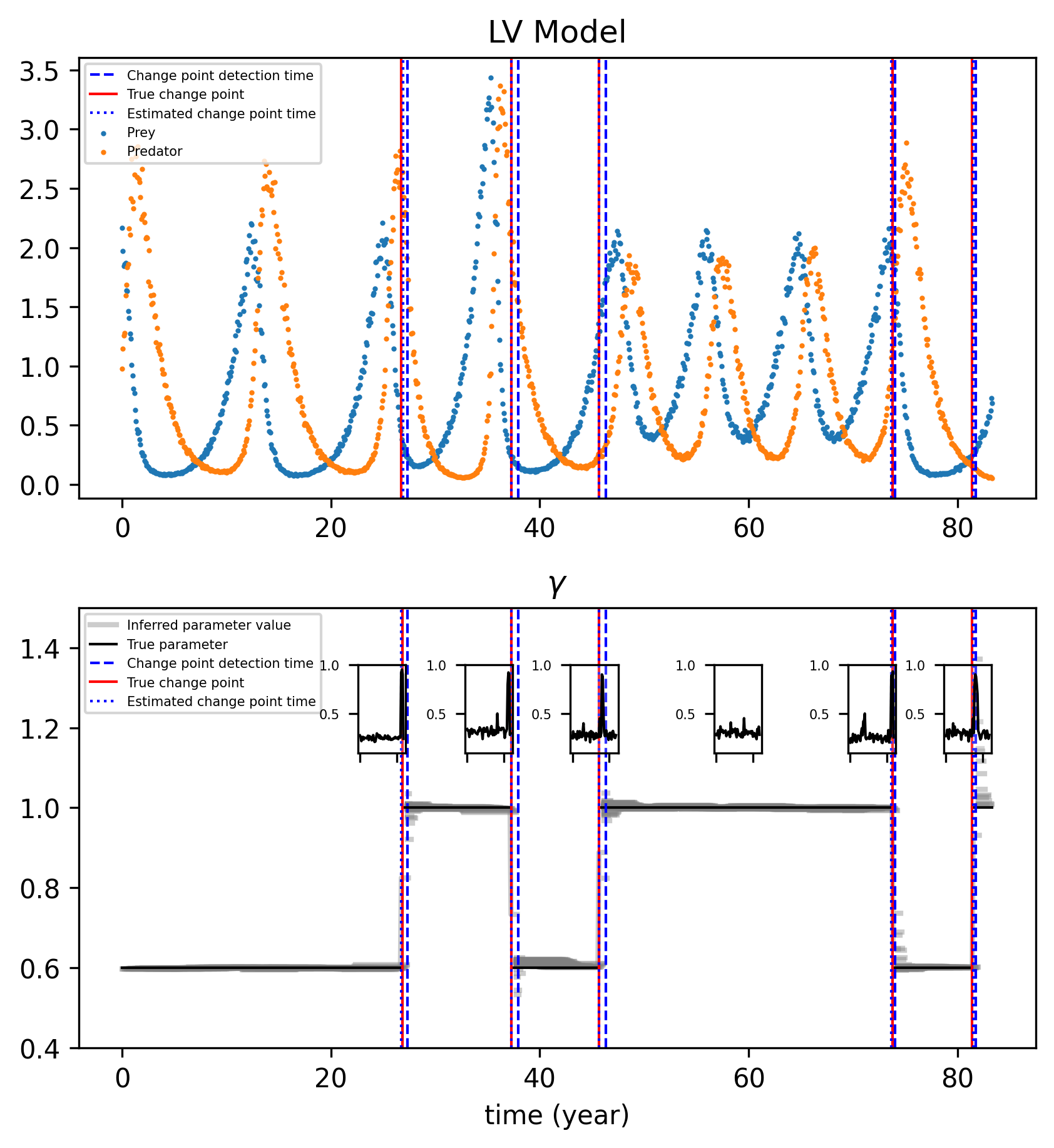}
     \end{subfigure}
\caption{Results of LV model, with sample observations (upper) and inferred parameters (lower). The six subplots in the lower figure denotes denote the probability of change point existence during six given time frames. }\label{fig:lv}
\end{figure}

The results of LV model are shown in Figure \ref{fig:lv}. Compared to SEIRD model observations, the changes are more difficult to identify through human inspection. Nonetheless, O-MAGIC is able to capture such changes with a small detection delay. Moreover, the estimated parameter $\gamma(t)$ exhibits high accuracy, underscoring the model's proficiency even when handling more complex ODE models.

For uncertainty quantification, we set prior parameters $\sigma_0=0.01, [\gamma_{\min},\gamma_{\max}] =[0, 1], \lambda_0=5$. 1000 posterior samples of $\bm A$ are drawn after discarding the first 500 burn-in samples. From the subplots in Figure \ref{fig:lv}, there is a notable increase in the posterior probability of a change point in the neighborhood of the true time point, as contrasted with other timestamps.

\begin{table}[htp]
\begin{center}
\scalebox{0.7}{
\begin{tabular}{ccccccc}
\toprule
Model name & FAR (\%)  & EDD & MSE & Cover & Computation time (h) \\
\midrule
O-MAGIC & 3.92 $\pm$ 2.77  & 5.42 $\pm$ 1.86 & 1.29 $\pm$ 0.51 & 0.97 $\pm$ 0.02 & 28.7 $\pm$ 0.3 \\
Runge-Kutta & 6.81 $\pm$ 3.98  & 4.62 $\pm$ 1.27 & 1.33 $\pm$ 0.54 & 0.95 $\pm$ 0.08 & 40.6 $\pm$ 0.5 \\
Microsoft SSA & 21.52 $\pm$ 11.30 & 13.51 $\pm$ 9.94 & 13.51 $\pm$ 9.94 & 0.68 $\pm$ 0.10 & 0.1 $\pm$ 0.0 \\
TIRE & 44.27 $\pm$ 15.39 & 11.19 $\pm$ 8.56 &  11.19 $\pm$ 8.56  & 0.61 $\pm$ 0.13 & 0.1 $\pm$ 0.0 \\
\bottomrule
\end{tabular}
}
\end{center}
\caption{The mean of each evaluation metric and computation time is reported first with the standard deviation across 100 replications followed after $\pm$ sign.}
\label{tab:lv}
\end{table}

\subsection{Lorenz model}
\begin{figure}[htp]
     \centering
     \begin{subfigure}[b]{1\textwidth}
         \centering
         \includegraphics[width=4in]{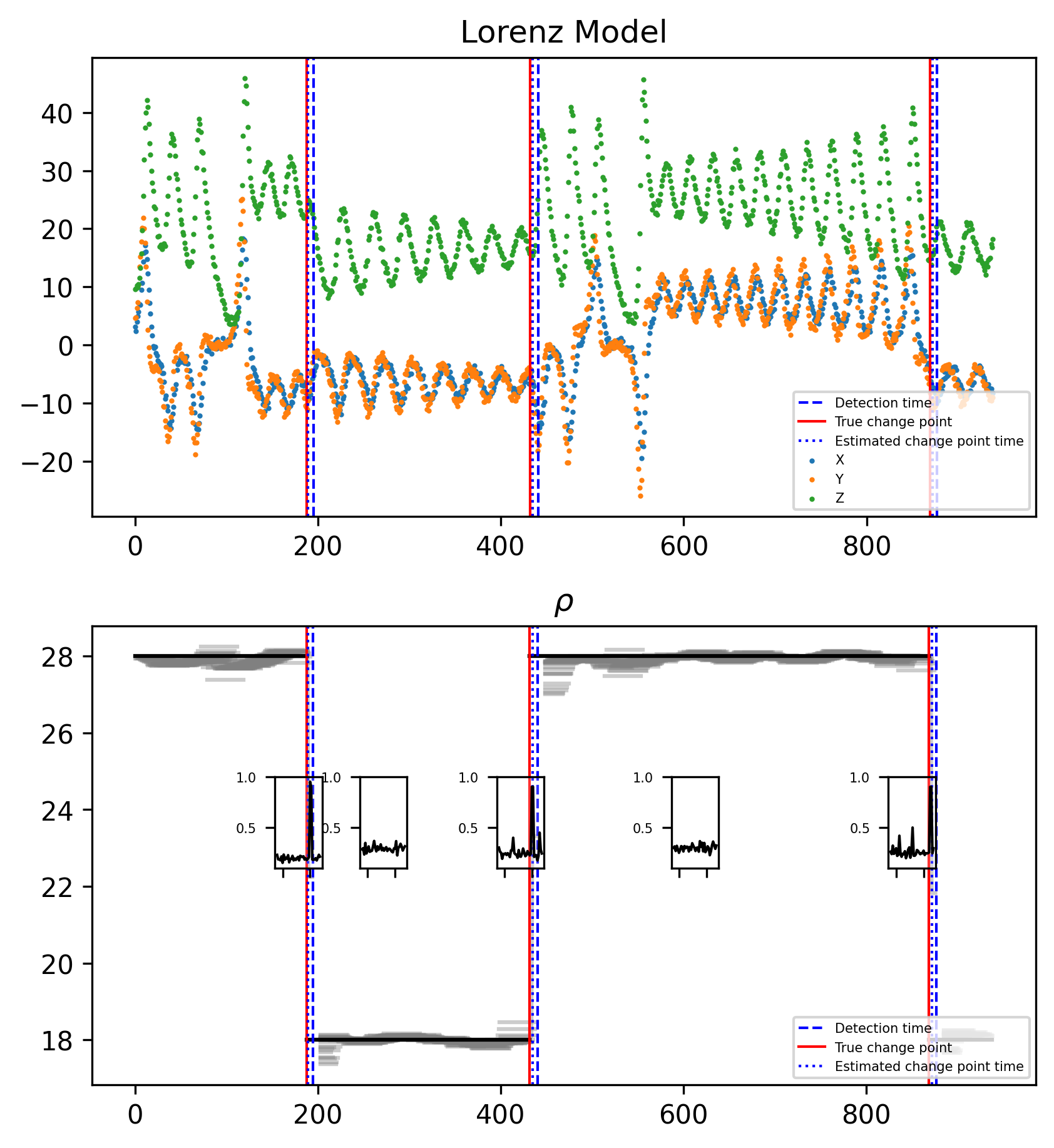}
     \end{subfigure}
\caption{Results of Lorenz model, with sample observations (upper) and inferred parameters (lower). }\label{fig:lorenz}
\end{figure}

In this section, we demonstrate the effectiveness of our approach in detecting state changes within chaotic systems. During periods of chaos, even though the observation patterns may show apparent discontinuity, the parameters might remain constant. This scenario often leads to the failure of many time-series-based methods when making inferences from chaotic observations. The Lorenz model, widely studied in chaotic dynamic systems, was first explored by mathematician and meteorologist Edward Lorenz \cite{yu2009dynamic}. This model is noted for its chaotic solutions under specific parameter values and initial conditions - particularly when $\rho$ is sufficiently large. Under these conditions, the future trajectory of the system is fundamentally unpredictable \cite{manneville1979intermittency}, rendering inferences particularly challenging. This model underscores that physical systems, while completely deterministic, can still pose significant computational challenges. The ODE structure of the Lorenz system is defined as follows:
\begin{align*}
    \frac{dx}{dt}=\sigma(y-x) \\
    \frac{dy}{dt}=x(\rho-z)-y \\
    \frac{dz}{dt}=xy-\beta z 
\end{align*}
In the experiment, following the setting of \cite{tykierko2007using}, we set $\sigma=10$, $\beta=2.667$, and Lorenz system transfers between chaotic ($\rho=28$) and non-chaotic ($\rho=18$) states. The initial values are randomly sampled from a broad range: $x_0, y_0, z_0 \sim U[-50, 50]$, with additive noise of $\sigma=1$. Figure \ref{fig:lorenz} shows the sample observation and the inferred parameters. Table \ref{tab:lorenz-1} shows the result of the detected change point. 

\begin{table}[htp]
\begin{center}

\scalebox{0.7}{
\begin{tabular}{ccccccc}
\toprule
Model name & FAR (\%)  & EDD & MSE & Cover & Computation time (h) \\
\midrule
O-MAGIC & 5.13 $\pm$ 4.81 & 4.37 $\pm$ 0.94 & 1.12 $\pm$ 0.43 & 0.94 $\pm$ 0.03 & 33.8 $\pm$ 0.2 \\
Runge-Kutta & 12.89 $\pm$ 7.20 & 5.85 $\pm$ 2.28 & 1.86 $\pm$ 1.05 & 0.87 $\pm$ 0.07 & 45.9 $\pm$ 0.3 \\
Microsoft SSA & 33.27 $\pm$ 10.11 & 19.32 $\pm$ 10.50 & 19.32 $\pm$ 10.50 & 0.72 $\pm$ 0.13 & 0.1 $\pm$ 0.0 \\
TIRE & 26.62 $\pm$ 11.43 & 15.64 $\pm$ 11.60 &  15.64 $\pm$ 11.60 & 0.76 $\pm$ 0.21 & 0.1 $\pm$ 0.0 \\
\bottomrule
\end{tabular}
}
\end{center}
\caption{The mean of each evaluation metric and computation time is reported first with the standard deviation across 100 replications followed after $\pm$ sign.}
\label{tab:lorenz-1}
\end{table}

In chaotic scenarios, O-MAGIC notably surpasses the brute-force numerical approach and other methods in terms of the false alarm rate. This superior performance can be attributed primarily to the fact that time series-centric methods erroneously segment chaotic periods into multiple intervals, even when the parameters (latent states) remain unchanged. Another contributing factor is the challenges posed by optimization; due to the chaotic nature of the observations, optimizations reliant on numerical solvers are susceptible to becoming trapped in local minima. This leads to imprecise statistical inference and a higher false alarm rate.

The prior parameters for uncertainty quantification are set as: $\sigma_0=0.01, [\beta_{\min},\beta_{\max}] =[0, 1],$ $[\rho_{\min},\rho_{\max}]=[10, 30], \lambda_0=5$. 1000 posterior samples of $\bm A$ are drawn after discarding the first 500 burn-in samples. Figure \ref{fig:lorenz} exhibits an elevated posterior probability of change point occurrence around the true time points.

\section{Discussion}
\label{discussion}
In this paper, we introduce a Bayesian approach, O-MAGIC, for parameter change detection for multi-dimensional time series with ODE dynamic structure. 
O-MAGIC conducts a generalized likelihood ratio test through modeling latent states as Gaussian process constrained to have the derivative processes satisfy ODE dynamics. O-MAGIC is shown to be statistically principled. Its general applicability in multi-dimensional, periodic, and chaotic dynamic systems is illustrated accordingly. Results have shown that O-MAGIC yields accurate and robust detection of change points from noisy observations, and with the distinction of detection time and the estimated change time, the accuracy of the change point can be further enhanced. Moreover, O-MAGIC can yield reasonable estimates of system parameters, which is helpful for further evaluation of the changes. 

\subsection{Advantages of including dynamic structure}
O-MAGIC is more accurate and efficient than the benchmark methods because it fully utilizes the structural information of the data while addressing the challenges of the numerical integration method. Brute force numerical integration methods are known to be the gold standard for general ODE change detection problems. However, with the presence of a chaotic system, due to the optimization limitation, the false alarm rate will significantly increase. Other approaches, which are built upon time series analysis or deep learning methods, do not integrate dynamic information, and thus fail to capture the changes of system states (parameters) accurately. O-MAGIC is the only approach aside from the brute-force method that is theoretically sound, practically accurate, computation efficient, and generally applicable for the ODE inference problem when time-constant and time-varying parameters co-exist.

Furthermore, the experiments have shown that the algorithm is relatively robust to larger threshold $h$ in the likelihood ratio test, which is critical to the performance of the algorithma small value of $h$ will result in a higher false alarm rate. 

During the online detection of parameter changes, the detection results can be influenced by the choice of threshold $h$. Through all simulation experiments, we find that the results are very robust to the choice of threshold, as the likelihood ratio will drastically increase when a parameter change occurs. Generally speaking, a higher threshold will lead to a higher miss alarm rate and a lower false alarm rate. Table \ref{tab:threshold} summarizes the model performance under different levels of threshold $h$ in the LV model. The model can perform well within a large range of $h$, which makes the choice of threshold less of an issue compared to classic change point topics. 

On the computational time comparison, O-MAGIC has the notable advantage of reduced computation cost compared to alternative methods that involve numerical integrations. The save in computation cost will be more significant when the stream of data accumulates, e.g., long-time systems.


\begin{table}[htp]
\begin{center}
\scalebox{0.7}{
\begin{tabular}{ccccccc}
\toprule
Threshold $h$ & FAR (\%) & MAR & EDD & MSE & Cover \\
\midrule
10 & 11.63 $\pm$ 4.58  & 0 & 3.07 $\pm$ 0.98 & 1.43 $\pm$ 0.64 & 0.86 $\pm$ 0.12 \\
20 & 7.13 $\pm$ 3.41  & 0 & 3.95 $\pm$ 2.11 & 1.27 $\pm$ 0.55 & 0.92 $\pm$ 0.10 \\
30 & 3.92 $\pm$ 2.77  & 0 & 5.42 $\pm$ 1.86 & 1.29 $\pm$ 0.51 & 0.94 $\pm$ 0.03 \\
50 & 3.03 $\pm$ 3.28  & 6.36 $\pm$ 4.97 & 9.74 $\pm$ 2.33 & 1.82 $\pm$ 0.96 & 0.89 $\pm$ 0.04 \\
\bottomrule
\end{tabular}
}
\end{center}
\caption{Comparison among different thresholds of LV model based on 100 simulation datasets. The mean of each evaluation metric is reported first with the standard deviation across 100 replications followed after the $\pm$ sign. The last row shows the computing time (in minutes) needed to obtain point estimates from all methods.}
\label{tab:threshold}
\end{table}

\subsection{Prior of $\theta$}    
In section \ref{sec:4.2.1}, We assume an uninformative prior on $\bm{\theta}$, for the sake of generality. However, other types of priors can be incorporated when additional expert knowledge is accessible. Specifically, let $\mathcal{F}$ denote the space of all piecewise constant functions on $\mathbb{R}^P$. Then, $\pi(\bm{\theta})\propto 1, \forall \bm{\theta}\in \mathcal{F}$. This distinct sample space for $\bm{\theta}$ separates our formulation from the conventional manifold constraint Gaussian process inference for ODE parameters \cite{yang2021inference}.
    
\subsection{Increase the robustness of algorithm}
Several measures are taken to decrease the false alarm rate caused by the optimization procedure. First, the likelihood under the alternative hypothesis is calculated, and the initial values of $\bm x$, $\bm \theta$, etc. from the null hypothesis are taken from the corresponding optimized values from the alternative hypothesis as a warm start. Second, a proper upper and lower bound of the kernel hyperparameters in Eq. \ref{hyper-update}, are helpful to prevent kernel over-fitting. Besides, parameters $\bm \psi$ and $\bm \sigma$ are updated within a small neighborhood during each optimization procedure to prevent too many updates in one single iteration. Lastly, the assumption of the minimal distance of adjacent change points greatly helps to reduce the false alarm rate. This is because, in our algorithm, the optimization may become less stable when the window size is small, resulting in an increased false alarm. Another advantage is the saving of computation costs as some iterations of GLR tests can be omitted. 

\subsection{Sensitivity analysis on uncertainty quantification prior}
\begin{figure}[htp]
     \centering
     \begin{subfigure}[b]{1\textwidth}
         \centering
         \includegraphics[width=4in]{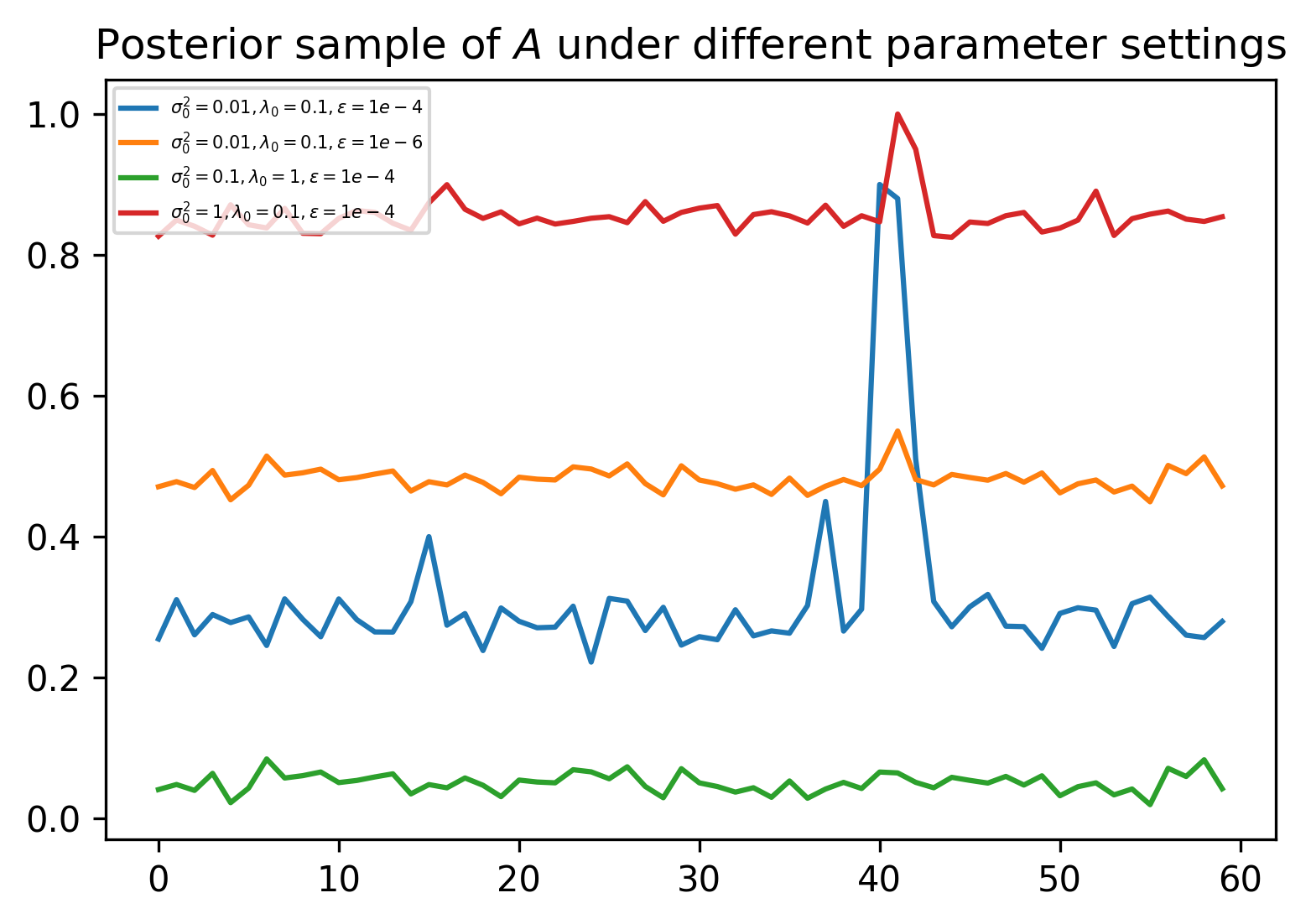}
     \end{subfigure}
\caption{Mean of posterior sample $\bm A$, under different parameter settings. True change occurs at $t=40$.}\label{fig:sensitivity}
\end{figure}

When estimating uncertainty, two types of parameters are involved: parameters that harness the prior of $\bm A$ (see Section \ref{sec:uncertainty}), including $\sigma^2_0, \lambda_0$, and parameters for HMC sampling: $\epsilon, N_{\text{step}}$. In this section, we discuss how these parameters influence the results of uncertainty quantification. Figure \ref{fig:sensitivity} shows the mean of the posterior sample for $\bm A$, under different parameter settings.

When $\sigma^2_0$ increases or $\lambda_0$ decreases, the prior changes will increase. Consequently, the overall probability of change increases for each time point. On the other hand, increase in $\sigma^2_0$ or decrease in $\lambda_0$ results in decreased mean probability. With a properly imposed prior of change points, most parameter settings can exhibit an evident signal of change point in the posterior samples. 

Moreover, the sampling related hyperparameters may also influence the performance of uncertainty quantification. When the step size $\epsilon_0$ is too large, the new samples tend to be always rejected, and the posterior samples cannot capture the parameter changes. When $\epsilon_0$ is too small, the proposed samples in each iteration will almost always be accepted, resulting in highly unstable samples, which weakens the signal of change. In practice, following the recommendations of \cite{girolami2011riemann}, hyperparameters shall be carefully tuned so that the HMC sampler has an overall acceptance rate of 70\%. 

There are several interesting future directions for O-MAGIC. We currently focus on the empirical performance of O-MAGIC through simulation examples. More theoretical studies on the identifiability issue and asymptotic optimality of the change-point estimation are all natural directions of future research. It would be also of future interest to extend O-MAGIC for change point detection in spatial-temporal dynamics \cite{chen2020spatial}, or stochastic differential equation of inherent noise modeling \cite{kou2004generalized}.


\bibliographystyle{unsrt}

\bibliography{Bibliography-MM-MC}
\newpage
\begin{center}
{\large\bf SUPPLEMENTARY MATERIAL}
\end{center}

\begin{algorithm} \label{alg1}

\caption{Efficient M-H algorithm for updating $\bm A$}\label{alg:hmc}
\textbf{Pre-set}: \\
$\sigma^2_0$: variance of random walk \\
$[\bm \theta_{\min}, \bm \theta_{\max}]$: lower and upper bound of parameters \\
$\lambda_0$: estimated rate of change points  \\
$\epsilon$: step size of HMC sampler \\
$L$: number of leaf frog steps \\
$N$: number of samples  \\
\textbf{Initialize}: \\
$\bm A^{cur}$: initial partition of time series as [1,0,0,...,0] \\
$\bm \theta^{cur}$: initial value of estimated $\theta$, given partition $\bm A^{cur}$ \\
$\bm x^{cur}$: initial value of estimated $\bm x$, given partition $\bm A^{cur}$ \\
$L^{cur}$: Posterior of $\bm x^{cur}$ and $\bm \theta^{cur}$, specified in Eq. \ref{cpd:posterior} \\
\begin{algorithmic}
\For{1:MaxIter}
    \For{i in $1 : N$}
    \State $\bm A^{new} \leftarrow \bm A^{cur}$
    \State $ A^{new}_i \leftarrow 1-A^{new}_i$ \Comment{Iteratively flip one bit of $\bm A$}
    \State Use HMC sampler   to sample $\bm \theta^{new}$ and $\bm x^{new}$ given $\bm A^{new}$, according to Eq. \ref{cpd:posterior} 
    \If {$\frac{P^{new}(\theta^{new}, \bm x^{new}, \bm A^{new})}{P^{cur}(\theta^{cur}, \bm x^{cur}, \bm A^{cur})}> U$}
    \Comment{Accept and update likelihood of each choice}
    \State $\bm A^{cur} \leftarrow \bm A^{new}$
    \State $\bm \theta^{cur} \leftarrow \bm \theta^{new}$
    \State $\bm x^{cur} \leftarrow \bm x^{new}$
    \State $\bm L^{cur} \leftarrow \bm L^{new}$
    \EndIf
    \EndFor
\EndFor
\end{algorithmic}
\end{algorithm}
\end{document}